\documentclass[aps,prb,reprint,showpacs,superscriptaddress,groupedaddress]{revtex4-1}  
\usepackage{dcolumn}   
\usepackage{bm}        
\usepackage{amssymb}   
\usepackage{amsfonts}  
\usepackage{amsmath}   
\usepackage{mathcomp}  
\usepackage{graphicx}  
\usepackage{subfigure} 
\usepackage{multirow}  
\usepackage{hyperref}  
\usepackage{longtable} 

\usepackage{verbatim}
\usepackage{color}
\usepackage[normalem]{ulem}

\begin{document}

\title{Exchange-correlation energy for the $3D$ homogeneous electron gas at arbitrary temperature}

\author{Ethan W. Brown}
\email{brown122@illinois.edu}
\affiliation{Department of Physics, University of Illinois at Urbana-Champaign, 1110 W.\ Green St.\ , Urbana, IL  61801-3080, USA}
\affiliation{Lawrence Livermore National Lab, 7000 East Ave, L-415, Livermore CA 94550, USA}
\author{Jonathan L. DuBois}
\affiliation{Lawrence Livermore National Lab, 7000 East Ave, L-415, Livermore CA 94550, USA}
\author{Markus Holzmann}
\affiliation{LPTMC, UMR 7600 of CNRS, Université Pierre et Marie Curie, 75005 Paris, France}
\affiliation{Université Grenoble 1/CNRS, LPMMC, UMR 5493, B.P. 166, 38042 Grenoble, France}
\affiliation{European Theoretical Spectroscopy Facility, Grenoble, France}
\author{David M. Ceperley}
\affiliation{Department of Physics, University of Illinois at Urbana-Champaign, 1110 W.\ Green St.\ , Urbana, IL  61801-3080, USA}
\date{\today}

\begin{abstract}
  We fit finite-temperature path integral Monte Carlo calculations of the exchange-correlation energy of the $3D$ finite-temperature homogeneous electron gas in the warm-dense regime ($r_{s} \equiv (3/4\pi n)^{1/3} a_{B}^{-1} < 40$ and $\Theta \equiv T/T_{F} > 0.0625$). In doing so, we construct a Pad\'{e} approximant which collapses to Debye-H\"{u}ckel theory in the high-temperature, low-density limit. Likewise, the zero-temperature limit matches the numerical results of ground-state quantum Monte Carlo, as well as analytical results in the high-density limit.
\end{abstract}

\pacs{}
\maketitle

\section{Introduction}
Density functional theory (DFT) is used ubiquitously in computational chemistry and condensed-matter physics \cite{RevModPhys.71.1253,RevModPhys.80.3}.   Recently there has been intense interest in extending the success of ground-state DFT to finite-temperature systems such as stellar, planetary interiors and other hot dense plasmas \cite{0953-8984-14-40-307,0741-3335-47-12B-S31,CTPP:CTPP239}. However, such attempts have met both fundamental and technical barriers when electrons have significant correlations.

There are two broad approaches to building finite-temperature functionals. In one approach, the exact Mermin finite-T DFT is approximated by smearing the electronic density of states over a Fermi-Dirac distribution \cite{PhysRev.137.A1441}. Although a useful approximation, this approach is not exact even in the limit of the exact ground state exchange functional as the Kohn-Sham orbitals need have no relation to the true excited states \cite{Karasiev20122519}. Additionally, as temperature increases, an ever-increasing number of molecular (Kohn-Sham) orbitals is required in order to evaluate the functional. This inevitably results in the DFT calculations becoming computationally intractable at some temperature. A second approach is to use Orbital-Free Density Functional Theory (OFDFT) where the usual Kohn-Sham orbitals are replaced by explicit density functionals for the kinetic energy and entropy terms \cite{PhysRevB.85.045125,PhysRevB.84.075146}. However, an a priori way to determine such functionals has yet to materialize. Without a reliable benchmark, OFDFT has historically been left to rely on Thomas-Fermi-like approximations which can incur errors an order of magnitude larger than typical DFT errors \cite{Karasiev20122519}. Recently generalized gradient approximations have improved OFDFT, introducing higher accuracy orbital-free kinetic energy density functionals for both 0-T and finite-T \cite{PhysRevB.81.045206,PhysRevB.86.115101}, as well as an exchange-correlation density functional for 0-T \cite{PhysRevLett.106.186406}. Nevertheless, the field still lacks a high-accuracy, orbital-free exchange-correlation energy density functional for finite-T.

In a recent paper, we provided accurate, first-principles thermodynamic data of the $3D$ homogeneous electron gas (HEG) throughout the warm-dense regime, making firm connections to both previous semi-classical and ground-state studies \cite{PhysRevLett.110.146405}. In that work we utilized the Restricted Path Integral Monte Carlo (RPIMC) method \citep{RevModPhys.67.279,binder1996monte,springerlink:10.1007/BF01030009}. Now, we fit this data to a functional form for the exchange-correlation energy which obeys the exact limiting behavior in temperature and density.

\section{Asymptotic Limits}
A satisfactory fit must match with known asymptotic limits. For the $3D$ HEG, analytic limits exist at high-temperature and low-density (the Debye-H\"{u}ckel limit), and at zero-temperature.

In the Debye-H\"{u}ckel (DH) limit, the quantum-mechanical Fermi-Dirac distribution may be approximated by the classical Boltzmann distribution, i.e. when $\Gamma \equiv 2/(r_{s} T) \ll 1$, where $T$ is in Rydbergs and $r_{s}$ is the Wigner-Seitz radius normalized by the Bohr radius. In this regime, the average potential energy per particle is much smaller than the thermal energy per particle, and each electron may be treated with a short-ranged, spherically-symmetric, screened interaction \cite{Debye:1923}. These approximations combined give the excess energy per particle to be $U_{DH} \equiv U-U_{0} = -\frac{\sqrt{3}}{2}\Gamma^{3/2}T = -\sqrt{6} r_{s}^{-3/2}T^{-1/2}$, where $U_{0}$ is the energy of an ideal gas (classically) or of a free Fermi gas (quantum mechanically). Classical simulations have numerically extended these results to larger values of $\Gamma$\cite{PhysRevA.8.3096,PhysRevA.8.3110}.

The first order quantum mechanical correction to these results is given through the Wigner-Kirkwood expansion in powers of $\hbar$, $U_{Q} = -\frac{\Gamma^{3}}{8}T^{2} = -r_{s}^{-3}T^{-1}$. The next order correction as well as the first-order exchange correction have also been calculated explicitly \cite{1975PhLA...53..187H,1978PhyA...91..152J}. Finally there has been some effort to calculate virial expansions of the excess energy at low-density and finite-temperature \cite{PhysRevE.53.5714}.

At zero-temperature, a significant body of numerical and analytical work has defined the exchange-correlation energy at all densities. In the high-density limit ($r_{s} \ll 1$) the total energy can be expressed as $E = a_{1}r_{s}^{-2} + a_{2}r_{s}^{-1} + a_{3}\log{r_{s}} + a_{4} + a_{5}r_{s}\log{r_{s}} + a_{6}r_{s} + \mathcal{O}(r_{s}^{2}\log{r_{s}})$. The first two coefficients can be determined through Hartree-Fock theory, with the first being the energy of a free Fermi gas and the second being the Fock exchange energy. Terms $a_{3}$ and $a_{4}$ were calculated by Gell-Mann and Brueckner \cite{PhysRev.106.364} using the random phase approximation (RPA). These results were extended by Carr and Maradudin \cite{PhysRev.133.A371} to determine $a_{5}$ and $a_{6}$. In the low-density limit ($r_{s} \gg 1$), one expects a body-centered cubic configuration, i.e. the Wigner crystal \cite{PhysRev.46.1002}. This suggests the form $E = A_{1}r_{s}^{-1} + A_{2}r_{s}^{-3/2} + A_{3}r_{s}^{-2} + A_{4}r_{s}^{-5/2} + \mathcal{O}(r_{s}^{-3})$ for the total energy. The first coefficient, the Madelung term, was first calculated by Fuchs \cite{1935RSPSA.151..585F}. The next three terms, coming from the zero-point harmonic vibration and its associated anharmonic corrections, were determined by Carr et al. \cite{PhysRev.124.747}.

High-precision quantum Monte Carlo (QMC) calculations have since spanned these two regimes \cite{PhysRevB.18.3126,PhysRevLett.45.566}, paving the way for accurate parameterizations which leverage the foregoing limiting forms \cite{PhysRevB.23.5048,PhysRevB.45.13244,doi:10.1139/p80-159}. Such functionals have been integral to the development and expansion of the local density approximation (LDA) of zero-temperature DFT~\cite{PhysRev.140.A1133}.

\section{Prior Fits}
Several attempts have been made at extending the success of ground state DFT to finite-temperature and this has resulted in the creation of a number of finite-temperature parameterizations of the exchange-correlation energy \cite{PhysRevB.34.2097,1986JPSJ55.2278T,PhysRevB.62.16536,PhysRevE.87.032102} A basic approach is the random phase approximation (RPA), which is accurate in the low-density, high-temperature limit (where it reduces to DH) and the low-temperature, high-density limit, since these are both weakly interacting regimes. Its failure, however, is most apparent in its estimation of the equilibrium, radial distribution function $g(r)$ which becomes unphysically negative for stronger coupling \cite{PhysRevE.87.032102}.

\begin{figure*}
  \centering
  \begin{subfigure}
          \centering
          \includegraphics[width=0.3\textwidth]{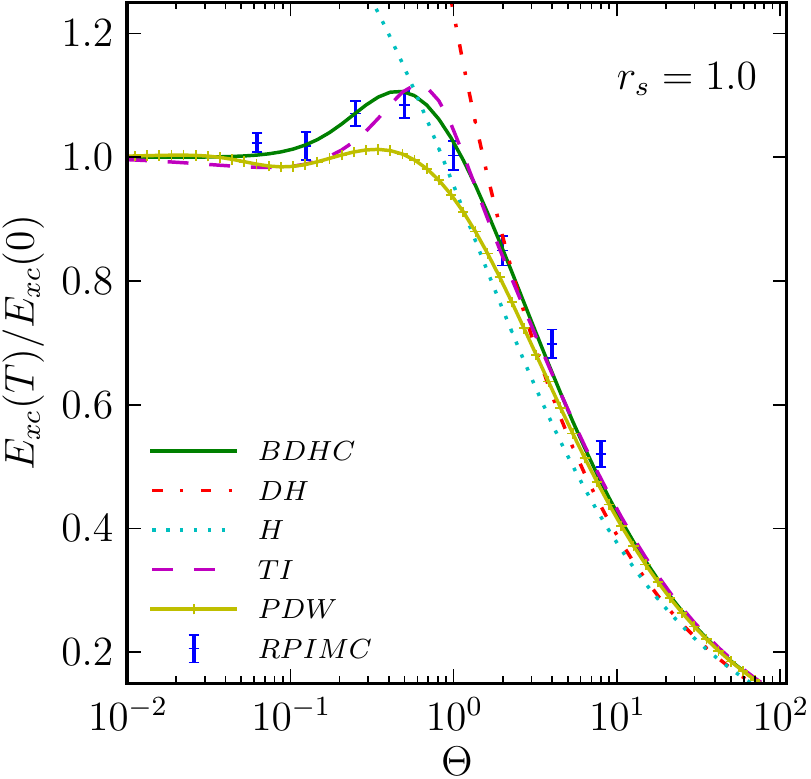}
  \end{subfigure}%
  ~ 
  \begin{subfigure}
          \centering
          \includegraphics[width=0.3\textwidth]{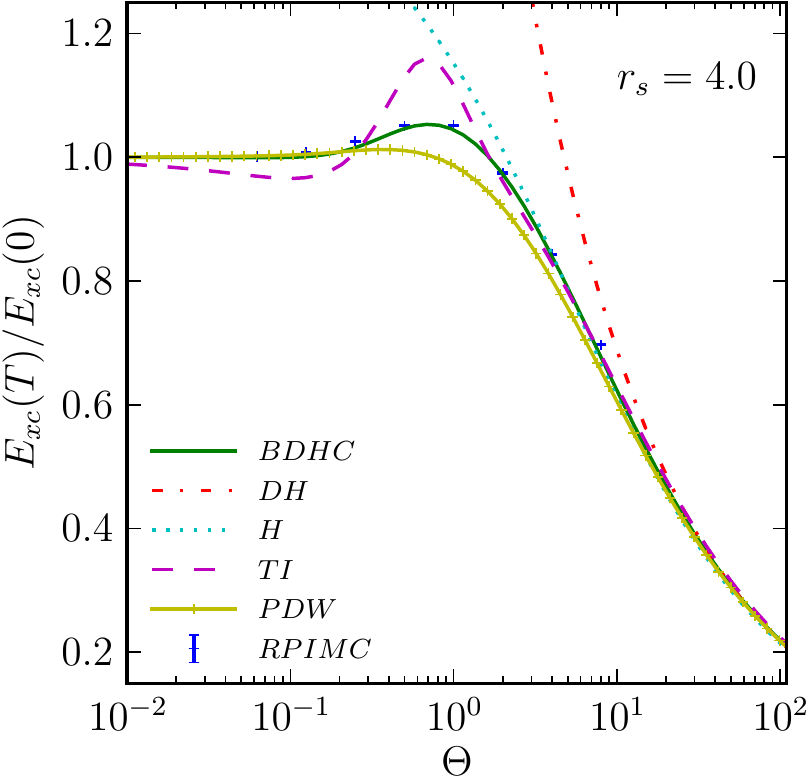}
  \end{subfigure}%
  ~ 
  \begin{subfigure}
          \centering
          \includegraphics[width=0.3\textwidth]{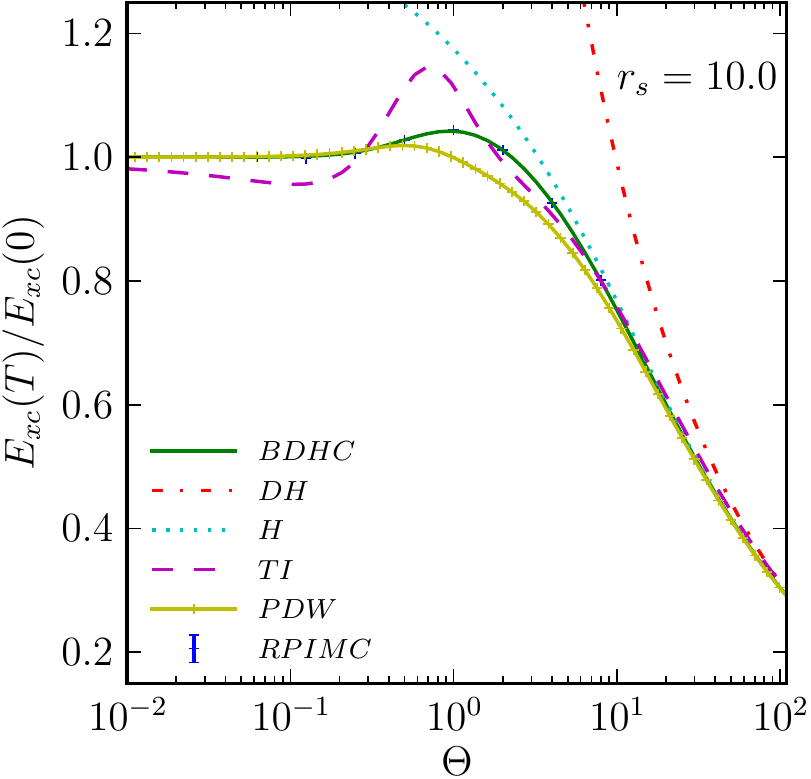}
  \end{subfigure}%
  \caption{(color online) Ratio of the exchange-correlation energy $E_{xc}$ at temperature $T$ to that at $T=0$ for the unpolarized $\xi=0$ $3D$ HEG with $r_{s}=1.0$, $4.0$, and $10.0$ (respectively). Shown are the results from numerical calculations (RPIMC), the present parameterization (BDHC), and several previous parameterizations. The latter include Debye-H\"{u}ckel (DH), Hansen (H), Tanaka and Ichimaru (TI), and Perrot and Dharma-wardana (PDW), all of which are discussed in the text.}
  \label{fig:SemiClassic-0}
\end{figure*}

\begin{figure*}
  \centering
  \begin{subfigure}
          \centering
          \includegraphics[width=0.3\textwidth]{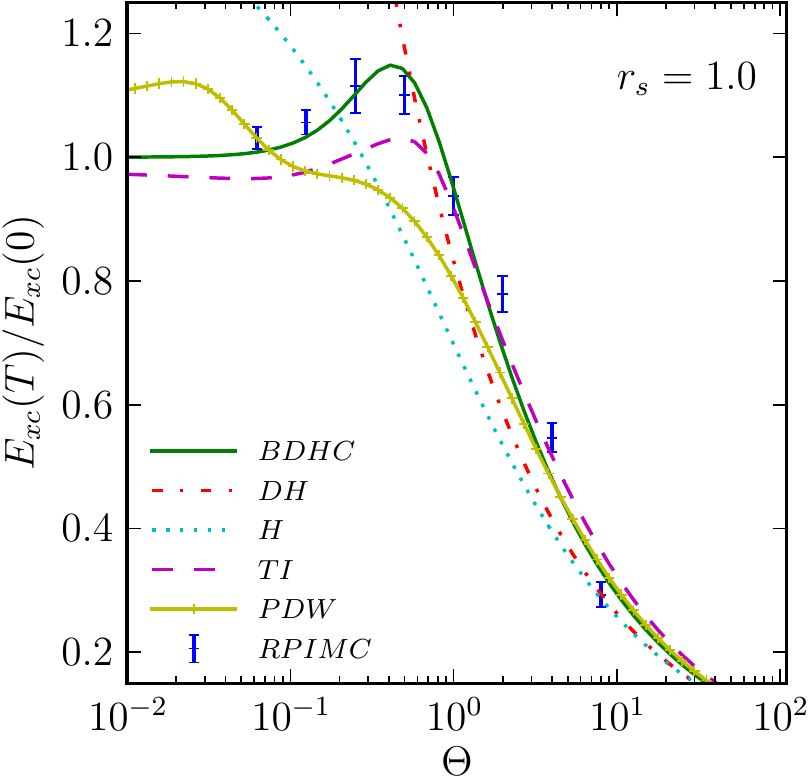}
  \end{subfigure}%
  ~ 
  \begin{subfigure}
          \centering
          \includegraphics[width=0.3\textwidth]{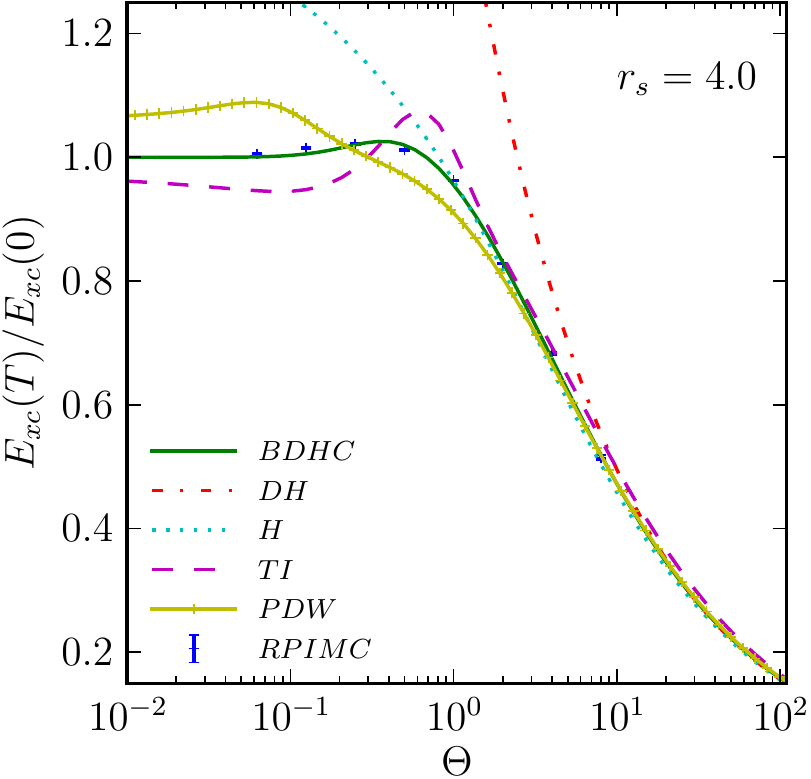}
  \end{subfigure}%
  ~ 
  \begin{subfigure}
          \centering
          \includegraphics[width=0.3\textwidth]{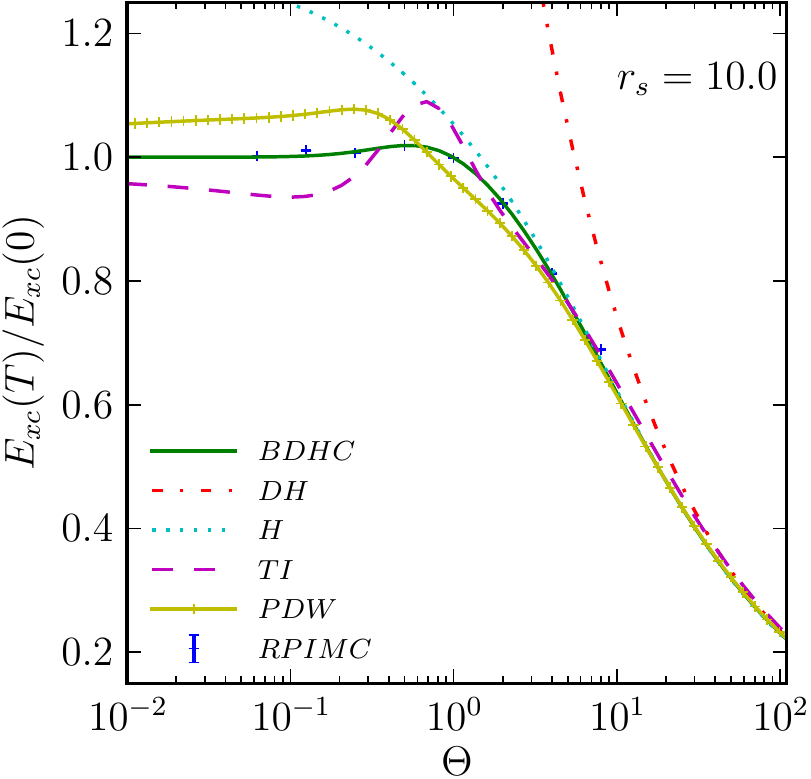}
  \end{subfigure}%
  \caption{(color online) Ratio of the exchange-correlation energy $E_{xc}$ at temperature $T$ to that at $T=0$ for the polarized $\xi=1$ $3D$ HEG with $r_{s}=1.0$, $4.0$, and $10.0$ (respectively). Shown are the results from numerical calculations (RPIMC), the present parameterization (BDHC), and several previous parameterizations. The latter include Debye-H\"{u}ckel (DH), Hansen (H), Tanaka and Ichimaru (TI), and Perrot and Dharma-wardana (PDW), all of which are discussed in the text.}
  \label{fig:SemiClassic-1}
\end{figure*}

Extensions of the RPA into intermediate densities and temperatures have largely focused on constructing local-field corrections (LFC) through interpolation since diagrammatic resummation techniques often become intractable in strongly-coupled regimes. Singwi, \emph{et. al.} \cite{PhysRev.176.589} introduced one such strategy relying on two assumptions. First, they use the static polarization-potential approximation allowing one to write the LFC, $G(k,\omega) \simeq G(k,\omega=0) \equiv G(k)$. Next they assume the two-particle distribution function is a function of the Fourier transformed momentum distribution, $n(r)$, and the pair-correlation function, $g(r)$, allowing a self-consistent solution for $G(k)$. Tanaka and Ichimaru \cite{1986JPSJ55.2278T} (TI) extended this method to finite temperatures and provided the parameterization of the $3D$ HEG correlation energy shown in Figs. \ref{fig:SemiClassic-0} and \ref{fig:SemiClassic-1}. A similar method by Dandrea \emph{et. al.} uses the Vashista-Singwi LFC \cite{PhysRevB.34.2097} to interpolate between the high- and low-temperature limits. Both methods appear to perform marginally better than the RPA at all temperatures, though both still fail to produce a positive-definite $g(r)$ at values of $r_s > 2$.

A third, more recent approach introduced by Perrot and Dharma-wardana (PDW) \cite{PhysRevB.62.16536} relies on a classical mapping wherein the distribution functions of a classical system at temperature $T_{cf}$, solved for through the hypernetted-chain equation, reproduce those for a quantum system at temperature $T$. In a previous work, PDW showed such a temperature $T_{q}$ existed for the classical system to reproduce the correlation energy of the quantum system at $T=0$ \cite{PhysRevLett.84.959}. To extend that work to finite temperature quantum systems, they use the simple interpolation formula $T_{cf} = \sqrt{T^{2} + T_{q}^{2}}$. This interpolation is clearly valid in the low-$T$ limit where Fermi liquid theory gives the quadratic dependence\cite{altland2010condensed} of the energy on $T$. Further in the high-$T$ regime, $T$ dominates over $T_{q}$ as the system becomes increasingly classical.

\section{Present Fit}
For our fit to RPIMC data, we employ a similar fitting functional as was used by PDW. To this end we define,

\begin{eqnarray}
  \label{eq:Exc}
  E_{xc}(r_{s},T) &\equiv& \frac{E_{xc}(r_{s},0) - P_{1}}{P_{2}}
\end{eqnarray}
where $E_{xc}(r_{s},0)$ is the ground-state exchange-correlation energy,
\begin{eqnarray}
  \label{eq:us}
  P_{1} &\equiv& (A_{2} u_{1} + A_{3} u_{2}) T^{2} + A_{2} u_{2} T^{5/2}, \\
  P_{2} &\equiv& 1 + A_{1} T^{2} + A_{3} T^{5/2} + A_{2} T^{3}, \\
  u_{1}(r_{s}) &\equiv& \frac{3}{2 r_{s}^{3}}, \\
  u_{2}(r_{s}) &\equiv& \frac{\sqrt{6}}{r_{s}^{3/2}},
\end{eqnarray}
and
\begin{eqnarray}
  \label{eq:As}
  A_{k}(r_{s}) &\equiv& \exp{[a_{k}\log{r_{s}} + b_{k} + c_{k} r_{s} + d_{k} r_{s} \log{r_{s}}]}.
\end{eqnarray}
Here $u_{1}$ and $u_{2}$ are chosen such that $\lim_{T\rightarrow \infty} E_{xc}(r_{s},T) = U_{DH} + U_{Q} + \mathcal{O}(T^{-3/2})$. The higher-order terms reflect the higher-order quantum corrections mentioned above. Likewise, note that $\lim_{T\rightarrow 0} E_{xc}(r_{s},T) = E_{xc}(r_{s},0) - \mathcal{O}(T^{2})$, reproducing both the ground-state exchange-correlation energy of Ceperley-Alder \cite{PhysRevLett.45.566} and the small-T quadratic behavior of Fermi liquid theory \footnote{Fitting exchange and correlation together avoids the cancelling $T^{2}\log{T}$ term coming from both.}. The Perdew-Zunger \cite{PhysRevB.23.5048} parametrization is used throughout for $E_{xc}(r_{s},0)$. The exchange-correlation energy between this and other parametrizations is at least two orders of magnitude smaller than the difference between the lowest temperature simulated and the Perdew-Zunger result. Because of this, we expect the use of another 0-T functional to have negligible effect on the finite-T parametrization we present.

We determine the best parameters of Eq. \ref{eq:As} through a least squares fitting of RPIMC data.\footnote{In fitting the data to this functional, it was noticed that the leading order, temperature dependent finite-size correction for the very high temperature points at small $r_{s}$ was not adequate. Instead, a more useful correction for these points extends from the classical regime. Again we may write the potential energy as $V = \frac{1}{2\Omega} \sum_{k} \frac{4\pi q^{2}}{k^{2}} S(k)$ where the structure factor is given by $S(k) = \frac{k^{2}}{m \omega_{p}(k)}[\frac{1}{\exp^{\omega_{p}/T} - 1} + \frac{1}{2}]$. Here $\omega_{p}^{2}(k) \equiv \frac{4\pi n e^{2}}{m} (1+k^{2}/k_{s}^{2})$ with $k_{s}^{2} \equiv 4\pi e^{2}/(\partial{\mu}/\partial{n})_{T}$, though since we are mostly concerned with the small $k$ limit, we take $\omega_{p}(k) \simeq \omega_{p} = 4\pi n e^{2}/m$. The finite-size correction then just reads as $\Delta V = V_{\infty}-V_{N}$. This correction is dominated by the long-wavelength ($k \rightarrow 0$) contribution. For $T \ll 1$ we recover the correction used in Ref. \cite{PhysRevLett.45.566}. For $r_{s} \gg T^{-2/3}$, however, we arrive upon $\Delta V = T/(2N)$ and thus $\Delta E = T/(4N)$. Through the virial theorem we then find $\Delta K = -T/(4N)$. This new correction was applied only to the points $r_{s} = 1.0, \Theta = 4.0,8.0$ and $r_{s} = 2.0, \Theta = 8.0$ for both the unpolarized and polarized system.} The RPIMC data shows a qualitative change in behavior around $r_{s} \approx 10$ and so we divide the fitting regime into two parts, $r_{s} < 10$ and $r_{s} > 10$. At $r_{s} = 10$, we make sure both the functional and its derivative are continuous. This is accomplished by ensuring each factor $A_{k}$ and its respective $r_{s}$ derivative is continuious at $r_{s} = 10$, providing 6 constraints and leaving 18 free parameters. For the unpolarized gas $\xi = 0$, we give the parameters in Table \ref{tab:Params-0}. Using these values, the fitting function has a maximum relative error of $0.9\%$. For the polarized gas $\xi = 0$, we give the parameters in Table \ref{tab:Params-1}. Using these values, the fitting function has a maximum relative error of $0.3\%$. Both of these maximum deviations occur at $r_{s} = 1.0$ where errors from RPIMC simulation were largest. All energies are in units of Rydbergs.

\begin{table}
  \caption{Fit parameters of the function in Eq. \ref{eq:As} for the unpolarized ($\xi = 0$) gas. The top table corresponds to $r_{s} < 10$, while the bottom table corresponds to $10 < r_{s}$.}
  \vspace{5 mm}
  \begin{tabular}{|c|c|c|c|c|}
    \hline
    $k$ & $a_{k}$ & $b_{k}$ & $c_{k}$ & $d_{k}$ \\
    \hline
    \hline
    $1$ & $3.56364$ & $-2.18158$ & $0.85073$ & $-0.28255$ \\
    \hline
    $2$ & $4.97820$ & $-2.72627$ & $0.62562$ & $-0.22889$ \\
    \hline
    $3$ & $9.41995$ & $-3.78699$ & $-1.87662$ & $0.39992$ \\
    \hline
    \hline
    $1$ & $4.38637$ & $1.22928$ & $-0.789404$ & $0.178368$ \\
    \hline
    $2$ & $5.96304$ & $0.249599$ & $-0.991637$ & $0.220769$ \\
    \hline
    $3$ & $5.43786$ & $-1.10198$ & $-0.716191$ & $0.157061$ \\
    \hline
  \end{tabular}
  \label{tab:Params-0}
\end{table}

\begin{table}
  \caption{Fit parameters of the function in Eq. \ref{eq:As} for the polarized ($\xi = 1$) gas. The top table corresponds to $r_{s} < 10$, while the bottom table corresponds to $10 < r_{s}$.}
  \vspace{5 mm}
  \begin{tabular}{|c|c|c|c|c|}
    \hline
    $k$ & $a_{k}$ & $b_{k}$ & $c_{k}$ & $d_{k}$ \\
    \hline
    \hline
    $1$ & $-1.57839$ & $-9.99823$ & $7.10336$ & $-2.19297$ \\
    \hline
    $2$ & $-1.46754$ & $-11.3387$ & $7.85547$ & $-2.40187$ \\
    \hline
    $3$ & $-0.784554$ & $-11.5341$ & $7.07407$ & $-2.17553$ \\
    \hline
    \hline
    $1$ & $-7.23836$ & $19.8258$ & $0.254584$ & $0.0521708$ \\
    \hline
    $2$ & $-6.65715$ & $19.9802$ & $0.263629$ & $0.0540244$ \\
    \hline
    $3$ & $-5.89226$ & $17.3632$ & $0.238536$ & $0.0488823$ \\
    \hline
  \end{tabular}
  \label{tab:Params-1}
\end{table}

\section{Discussion and Conclusions}
In Figs. \ref{fig:SemiClassic-0} and \ref{fig:SemiClassic-1}, we plot our fit, the RPIMC data, and all mentioned prior fits of the finite-temperature exchange-correlation energy. Clearly, the classical Debye-H\"{u}ckel limit is obeyed by each fit. However, only our fit and PDW obey the correct zero-temperature behavior ($E_{xc}(\Theta)/E_{xc}(0) \rightarrow 1$ as $\Theta \rightarrow 0$). The STLS driven fit of Tanaka and Ichimaru (TI) only agrees well with the RPIMC data at high-density -- i.e. where the RPA, the basis of STLS, is most applicable.

The PDW line in Figs. \ref{fig:SemiClassic-0} and \ref{fig:SemiClassic-1} clearly matches well with the RPIMC results in both temperature limits. It is not surprising, however, that in the intermediate temperature regime, where correlation effects are greatest, the quadratic interpolation of the temperature fails. A similar approach by Dutta and Dufty \cite{PhysRevE.87.032102} uses the same classical mapping as PDW, matching the $T=0$ pair correlation function instead of the correlation energy. While this gives accurate results near $T=0$, the breakdown of Fermi liquid behavior near the Fermi temperature causes the method to overestimate the exchange hole of the pair correlation function. A direct comparison of $E_{xc}$ is not yet available.

Finally we note that there has been some previous work on the low-density phases of $3D$ HEG both at $T=0$ \cite{2002PhRvE66c6703Z} and $T>0$ \cite{PhysRevLett.76.4572}. These include a predicted second-order transition to a partially polarized state around $r_{s} \simeq 50$, and a first-order transition into a Wigner-crystal for $r_{s} > 100$. Since both these transitions are outside the range of the fit data, we do not expect to see these transitions with the above functional.

In summary we have performed a least squares fitting of recent RPIMC data to a functional form which reproduces both high- and low-temperature asymptotic limits exactly. This fit outperforms all previous attempts at parameterizing the exchange-correlation energy at arbitrary temperature. We are providing a simple script of the functional in the Supplementary Material as well as at \url{http://github.com/3dheg/BDHC}. It is our hope that this newly created parameterization will be useful as a basis for new finite temperature DFT functionals and as a benchmark for orbital-free DFT studies.

\section{Acknowledgments}
The authors would like to thank Jeremy McMinis and Miguel Morales for useful discussions. This work was supported by grant DE-FG52-09NA29456. In addition, the work of E. Brown and J. DuBois was performed under the auspices of the U.S. Department of Energy by Lawrence Livermore National Laboratory under Contract DE-AC52-07NA27344 with support from LDRD 10-ERD-058 and the Lawrence Scholar program. Computational resources included LC machines at Lawrence Livermore National Laboratory through the institutional computation grand challenge program.

\bibliography{rapidcomm}{}


\end{document}